\documentclass[pre,aps,twocolumn,floatfix,superscriptaddress]{revtex4-1}
\usepackage{eso-pic,calc}
\usepackage{graphicx}
\usepackage{amsmath, amsthm, amssymb}
\usepackage{epsfig}
\usepackage{bm}
\usepackage{color}

\usepackage[colorlinks=True]{hyperref}  
\hypersetup{
 	colorlinks=true,  
 	linkcolor=blue,  
 	citecolor=blue, 
 	filecolor=magenta,   
 	urlcolor=blue         
 }

\def\ie{i.e.,~}
\def\etl{$et~al.$~}

\begin{document}
\title{Stationary $1/f^{\alpha}$ noise in discrete models of the Kardar-Parisi-Zhang class}

\author{Rahul Chhimpa}
\affiliation{Department of Physics, Institute of Science,  Banaras Hindu University, Varanasi 221 005, India}

\author{Avinash Chand Yadav\footnote{jnu.avinash@gmail.com}}
\affiliation{Department of Physics, Institute of Science,  Banaras Hindu University, Varanasi 221 005, India}

\begin{abstract}
{In discrete models describing growing rough interfaces of the Kardar-Parisi-Zhang universality class, we examine height fluctuations at a fixed site as a function of time in the monolayer unit. For small systems, we show that it is possible to reach the stationary state. We compute the two-time autocorrelation and power spectra independently. The correlation function remains non-exponential and vanishes after a correlation time that diverges with system size. As a result, the power spectra display a lower cutoff that maintains constant power. In the nontrivial frequency regime, we observe $1/f^{\alpha}$-type scaling with the spectral exponent 5/3. Finite-size scaling reveals that the temporal correlation function follows a dynamic scaling. Our findings, supported by scaling-theoretical arguments, establish that the fluctuations are wide-sense stationary, implying applicability of the Wiener-Khinchin theorem.}
\end{abstract}
\maketitle

\section{Introduction}
Astonishingly diverse non-equilibrium systems exhibit the emergence of long-range correlation. In the dynamics context, one can recognize such behavior as $1/f$ noise~\cite{dutta_horn_1981}, a scaling feature of the power spectral density (PSD) of a noisy signal. For a signal $X(t)$ measured up to a finite observation time $T$, the PSD is $S(f, T) = \langle |\tilde{X}(f, T)^2|\rangle/T$, where $\tilde{X}(f, T)$ is the truncated Fourier transform of the signal and $\langle \cdot \rangle$ denotes ensemble average. While signals ranging from voltage across electronic devices~\cite{johnson_1925, pellegrini_1983} to biological activities~\cite{Li_1992, siwy_2002, Sarlis_2009, Aguilar2019, singh_2023} display the $1/f$ scaling behavior, no general explanation exists. Considering the stationary or non-stationary property, the study of $1/f$ noise belongs to two approaches~\cite{Takeuchi_2017}. (i) If the signal represents stationary or long-time dynamics, then $S(f) = \lim_{T\to \infty}S(f, T)$. The two-time autocorrelation function $C(t_1, t_2) = \langle X(t_1)X(t_2)\rangle$ for a real signal becomes a function of an argument that is the absolute value of the difference between two times, $C(\tau = |t_1-t_2|)$, because of time translation symmetry. For such wide-sense stationary processes, the Wiener-Khinchin theorem suggests that the PSD is the Fourier (cosine) transform of the correlation function~\cite{kubo_1995}. We emphasize that the underlying temporal correlation function in several systems exhibits nonexponential behavior and dynamical scaling in terms of correlation time due to finite-size effects~\cite{ising_2025}. Correspondingly, the PSD displays a signature of a lower cutoff frequency (an inverse of the correlation time) below which the power remains constant~\cite{chhimpa_2025_division, Chhimpa_2025_sandpile}.

(ii) In experimental and theoretical studies, the PSD $S(f, T)$ of many processes (such as unrestricted Brownian motion and blinking quantum dots~\cite{Sadegh_2014}) lacks a lower cutoff frequency that remains elusive for a finite observation time. Such processes are basically nonstationary (can describe a transient towards a stationary state of an infinite-size system). The Wiener-Khinchin theorem becomes invalid in this context~\cite{kubo_1995, barkai_2015}. The PSD can (or does not) show dependence on the finite observation time. The two-time autocorrelation function does not satisfy the time-translational symmetry and remains a function of two times. Also, the ensemble average and time average are not the same. Recent works have emphasized that the Wiener-Khinchin theorem for nonstationary scale-invariant processes can provide the corresponding PSD if the correlation function satisfies dynamical scaling~\cite{barkai_2015, andreas_2015, leibovich_2016}.

In particular, we focus on kinetic rough interfaces that remain extensively studied nonequilibrium systems and exhibit interlinked space-time scaling features~\cite{Stanley_1995}. Among many relevant interfaces, the Kardar-Parisi-Zhang (KPZ) universality class~\cite{HalpinHealy2015, kazumasa_2018} represents an interesting phenomenon with diverse and growing applications found in isotropic spin chains~\cite{takeuchi_2025} and exciton-polariton condensates~\cite{Fontaine2022}. Experimentally, the study of such interfaces has been relevant to thin film deposition~\cite{PALASANTZAS2002357, almeida2014}, liquid crystal films~\cite {masaki_2010, Takeuchi2011}, growth of bacterial colonies~\cite{wakita_1997, MATSUSHITA1998517}, and cancer cell growth dynamics~\cite{huergo2012}. Numerical work relied on examining discrete models such as ballistic deposition (BD)~\cite{Meakin_1986, farnudi_2011} and the single-step (SS) model~\cite{Daryaei_2020}. Analytically, a continuum equation for the KPZ class~\cite{kpz_1986, Kriecherbauer_2010, Ivan_2012, fontaine_2023} describing the space-time evolution of the height profile $h(x, t)$ is 
\begin{equation}
\partial_t h = \nu \partial_{x}^{2} h  + \lambda \left(\partial_x h\right)^2 + \eta,
\label{eq_1}
\end{equation}
where $\nu$ is the surface tension of the interface and $\lambda$ is the strength of nonlinear term. The white Gaussian noise term $\eta$ has zero mean. The nonlinear term in Eq.~(\ref{eq_1}) accounts for the growth locally normal to the surface. The square of the width $w^2 (t, L) = \langle [h-\langle h\rangle]^2\rangle$ satisfies dynamic or Family–Vicsek scaling~\cite{Family_1985} as $w^2(t, L) \sim L^{2\chi}F_{w^2}(t/L^z)$, where the scaling function varies as $F_{w^2}(u) \sim u^{2\chi/z}$ for $u\ll 1$ (growth regime) and constant for $u\gg 1$ (saturation regime). $z$ and $\chi$ are the dynamic and roughness exponents, and the angular bracket $\langle \cdot \rangle$ denotes average over space and configurations.
As is known, the width fluctuations exhibit a saturation regime after a transient. The spatial correlation length varies as $\xi \sim L$. It also reveals the existence of correlation time $\mathcal{T} \sim L^{z}$.

In discrete models belonging to the KPZ universality class, Takeuchi~\cite{Takeuchi_2017} examined a temporal signal representing the interface's height fluctuations by computing the power spectra $S(f, T)$ for different values of the finite observation time $T$. He also supported the simulation results with the PSD characterization by experimentally studying growing interfaces in liquid-crystal turbulence. Interestingly, the height displays a signature of the $1/f^{\alpha}$ noise with a spectral exponent of 5/3. The power does not explicitly depend on the finite observation time. However, the power spectrum lacks a lower cutoff frequency, implying the underlying signals are non-stationary.

In the discrete KPZ model with large system size $L\to \infty$, Henkel \etl~\cite{henkel_2012} computed the two-time autocorrelation (ensemble-averaged) of the height noises. The correlation function reveals non-exponential behavior and dynamical scaling, but the time translational symmetry does not hold. These findings further support that the dynamics remain transient towards a stationary state.
If we assume an infinite-size system, then it is not possible to reach a stationary state in the finite observation time, implying that the dynamics of the non-equilibrium system would be non-stationary. The underlying two-time autocorrelation function shows an aging behavior. The typical aging behavior characterized by the two-time function exhibits the following characteristics: (i) A non-exponential relaxation, (ii) dynamic scaling, and (iii) lack of time-translational invariance~\cite{henkel_2012, Bao2024, Henkel_2025}.

In this contribution, we also consider discrete models belonging to the KPZ universality class. We reexamine the interface height fluctuations at a fixed site as a function of time in the monolayer unit. For a small system, we show that it is possible to reach the long-time limit or stationary state dynamics within a finite (long) observation time. We compute the two-time autocorrelation and power spectra independently. The correlation vanishes after a correlation time that diverges with system size. Moreover, the correlation function remains non-exponential and satisfies a dynamic scaling. Consequently, the power spectra exhibit a lower cutoff displaying constant power. In the non-trivial frequency regime, the power spectrum exhibits $1/f$-type scaling with the spectral exponent 5/3. Our findings, supported by scaling-theoretical arguments, establish that the dynamics are wide-sense stationary and the Wiener-Khinchin theorem is applicable.

The paper organization is as follows. In Sec.~\ref{sec_2}, we begin by recalling four discrete models describing rough interfaces of the KPZ universality class. We present numerical results for the temporal correlation and PSD of the height noises in the stationary state, along with a finite-size scaling (FSS) in Sec.~\ref{sec_3}. We also add a scaling argument here to support our results for the two characterizations in Sec.~\ref{sec_4}. Finally, the paper concludes with a discussion in Sec.~\ref{sec_5}.

\section{KPZ lattice models}{\label{sec_2}}
To illustrate the signature of stationary fluctuations for finite system size, we simulate the following discrete (cellular automata) models belonging to the KPZ universality class. Particularly, we examine the Kim-Kosterlitz (KK) model~\cite{kim1989}, the BD model~\cite{Meakin_1986}, the etching model (EM)~\cite{mello_2001}, and the SS model~\cite{Daryaei_2020}. We consider a one-dimensional lattice of size $L$ with periodic boundary conditions as a substrate. To each site, \ie $x \in [1, L]$, we associate a height variable $h(x, t)$ representing the number of deposited particles. Initially, the height variable $h$ is set to 0, indicating that the lattice is empty. A random column is selected to deposit a particle; the particle follows well-defined rules to stick to the surface, and the rules are defined below for each model. Since the particle doesn't always stick to a random site, it creates a nonuniform (rough) growth of the interface. We consider a monolayer time unit, \ie $L$ deposition attempts constitute one Monte Carlo time step.

In the KK model, at each randomly selected column, a Kim-Kosterlitz slope constraint $h(x,t)-h({x\pm1},t) \le 1$ is checked. The height of the selected column increases by one unit only if the constraint is satisfied; otherwise, we pick a new column similarly. The KK model was presented as a simple model to understand the essential features of the KPZ universality class~\cite{kim1989}.

The BD was initially introduced to model the colloidal aggregation. In this model, the particle falls vertically to the randomly selected column; however, it adheres to the earliest contact. The simulation can be easily performed by following a simple rule: If at time $t$ the height is $h(x, t)$ to a randomly selected column, then $h(x, t+1) = \max\{h(x\pm1, t), h(x, t)+1\}$~\cite{Meakin_1986}. The simple rule yields a nonlinear growth in the vertical direction. Interestingly, the model falls into the KPZ universality.

The third model we consider is the etching model. It was introduced as an atomistic model by Mellow \etl to study the etching of a solid~\cite{mello_2001}. We consider the deposition version of the EM. We increase the height of a randomly selected column $h(x, t)$ by one unit, \ie $h(x, t+1)=h(x, t)+1$, and the nearest neighbor columns modify as $h(x\pm1,t+1)=h(x, t)$.

The last model that we studied is the SS model. In the SS model, we start with a groove initial condition $h(x, 0)=[1+(-1)^x]/2$. If the randomly selected column is a local minima (maxima), then we add (remove) 2 particles with probability $p$ or $(1-p)$. It is interesting to note that the height difference between neighboring sites always remains 1. We set $p=1$ so the SS model belongs to the KPZ universality class~\cite{Daryaei_2020}.

\begin{figure}[t]
	\centering
	\scalebox{1.0}{\includegraphics{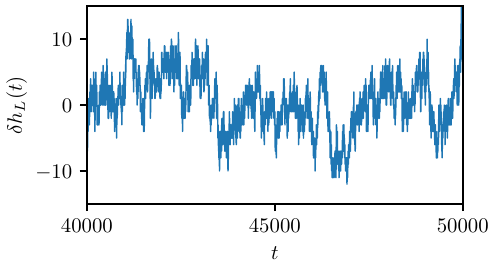}}
	\caption{A typical profile of the height noise signal $\delta h_L(t)$ in the KK model with the system size $L=2^8$. The signal $\delta h_L(t)$ represents the difference of the height at a site and the mean height of all the sites at time $t$. The time unit is considered monolayer, \ie one Monte Carlo time step is equivalent to $L$ deposition attempts.}
	\label{fig_signal}
\end{figure}

\begin{figure}[t]
	\centering
	\scalebox{1.0}{\includegraphics{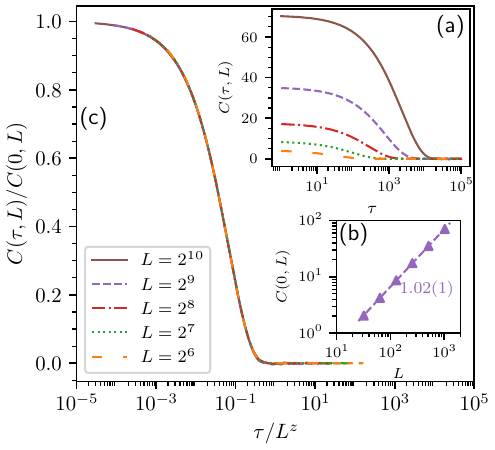}}
	\caption{(a) The two-time autocorrelation function $C(\tau, L)$ of the signal $\delta h_L(t)$ [cf. Eqs.~(\ref{corr_def}) and (\ref{corr_funct_1})]. The lag time $\tau$ ranges from 1 to $10^5$, and the signal length $T$ is $2^{22}$. We used 120 independent realizations of the process for the ensemble average. (b) As the system size increases, the zero-lag correlation $C(0, L)$ grows linearly. (c) The data collapse is obtained by plotting $C(\tau, L)/C(0, L)$ with $\tau/L^z$, where $z$ is the dynamic exponent.}
	\label{fig_kk_corr1}
\end{figure}

\section{results}{\label{sec_3}}
Let $h(x, t)$ be the height of the fluctuating rough interface at position $x$ after time $t$, which we measured in the monolayer units. We consider $L$ particles deposited as a unit time for a one-dimensional substrate of size $L$. As the mean height $\bar{h}(x, t) = \sum_{x =1}^L  h(x, t)/L$ (spatially averaged) grows with time typically linearly, we consider a well-behaved temporal signal as height fluctuations,
\begin{equation}
	\delta h_L(t) = h(x, t) - \bar{h}(x, t).
	\label{eq_h}
\end{equation}
To uncover the underlying temporal correlations of the signal described in Eq.~(\ref{eq_h}), we first compute the two-time autocorrelation function  
\begin{equation}
	C(\tau, L) = \langle \delta h_L(t)\delta h_L(t+\tau) \rangle,
	\label{eq_corr}
\end{equation}
where the angular bracket $\langle \cdot \rangle$ in Eq.~(\ref{eq_corr}) denotes average over independent realizations of the interfaces. Since the periodic boundary conditions ensure that the system remains homogeneous in space, the correlation function of the temporal fluctuations $\delta h_L(t)$ remains the same for all sites.

Numerically, starting from a flat initial condition $h(x, t=0) = 0$ and discarding transients up to $10^3$ for a process with system size $L = 2^{10}$, we record the interface height $h(x, t)$ at a fixed position $x$ with varying time $t$. We compute the height noise $\delta h_L(t)$ by removing the mean height (cf. Fig.~\ref{fig_signal}). We set the finite-observation time $T$ much larger than the underlying correlation time of the process that varies as $\mathcal{T} \sim L^z$, where $z$ is the dynamic exponent. It implies that the dynamics reached a stationary state. To determine the two-time autocorrelation function numerically, we follow the time average and the ensemble average, both as 
\begin{equation}
	C(\tau, L) = \langle (T-\tau)^{-1}\sum_{t=1}^{T-\tau} \delta h_L(t)\delta h_L(t+\tau) \rangle.
	\label{corr_def}
\end{equation}
Fig.~\ref{fig_kk_corr1}(a) shows the correlation function $C(\tau, L)$ for different system sizes $L = 2^6, 2^7, \cdots, 2^{10}$. As shown in Fig.~\ref{fig_kk_corr1}(b), the zero lag correlation or variance varies linearly with the system size $C(0, L) \sim L$. Finally, we plot the scaling function obtained by FSS as the variation of the normalized correlation function $C(\tau, L)/C(0, L)$ with the reduced lag time with respect to the correlation time $\tau/L^z$. The data collapse of the correlation function suggests that we can write a scaling ansatz,  
\begin{equation}
	C(\tau, L) = L F_C\left(\frac{\tau}{L^z}\right),
	\label{corr_scal_ansatz}
\end{equation}
where the scaling function, in the long time limit, demands a system size independent behavior for $\tau \ll L^z$ as
\begin{equation}
	F_C(\nu) \sim \begin{cases} \nu^{1/z},~~~~~~{\rm for}~ \nu \ll 1, \\ 0, ~~~~~~~~~~ {\rm for}~ \nu \gg 1. 
	  \end{cases}
	  \label{corr_scal_1}
\end{equation}
Eventually, we note that the correlation function varies as [cf. Eqs.~(\ref{corr_scal_ansatz})-(\ref{corr_scal_1})]
\begin{equation}
	C(\tau, L) \sim \begin{cases} \tau^{1/z},~~~~~~{\rm for}~ \tau \ll L^z, \\ 0, ~~~~~~~~~~ {\rm for}~ \tau \gg L^z.   \end{cases}
	  \label{corr_funct_1}
\end{equation}

\begin{figure}[t]
	\centering
	\scalebox{1.0}{\includegraphics{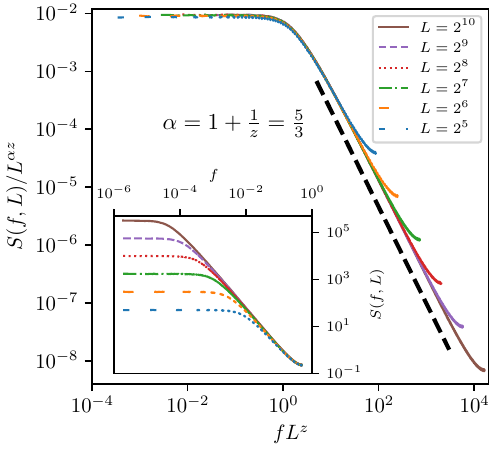}}
	\caption{Inset: The power spectrum $S(f, L)$ of the height noise $\delta h_L(t)$ in the KK model. The system size varies from $2^5$ to $2^{10}$. The initial $10^3$ measurements are discarded as transient, and the signal length is $T = 2^{20}$. To reduce noise in the PSD for each curve, we perform an ensemble average over $10^4$ independent realizations. Main panel: The data collapse of the power spectra as shown in the inset [cf. Eqs.~(\ref{eq_psd_1})-(\ref{eq_psd_3})]. The dashed (black) line guides the slope of 5/3.}
	\label{fig_psd_0}
\end{figure}

\begin{figure}[t]
	\centering
	\scalebox{1.0}{\includegraphics{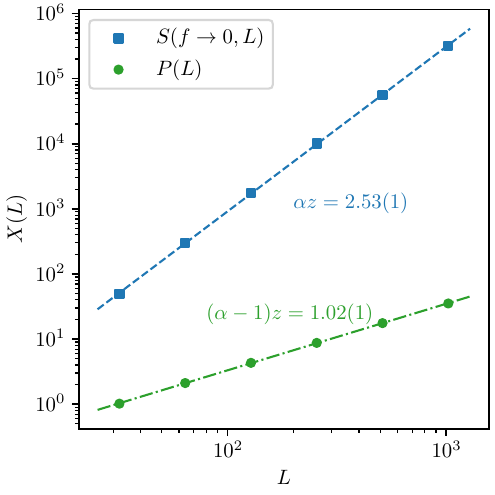}}
	\caption{The system size scaling of the power in the low frequency regime below the cutoff frequency $ \sim L^{-z}$. The lower curve shows the scaling of total power with system size. We numerically find the dynamic exponent $z = 1.51(2)$, which agrees well with the theoretical value 3/2.}
	\label{fig_psd_2}
\end{figure}

As the two-time autocorrelation function is a function of an argument that is the difference of the two times, the underlying signal is wide-sense stationary. We can apply the Wiener-Khinchin theorem to obtain the PSD by taking the Fourier transform of the correlation function, 
\begin{equation}
	S(f, L) \sim \int C(\tau, L) \exp{(-i2\pi f\tau)} d\tau.
	\label{eq_ft}
\end{equation}
Plugging Eq.~(\ref{corr_scal_ansatz}) into Eq.~(\ref{eq_ft}), we can write
\begin{equation}
	\nonumber
	S(f, L) \sim L^{z+1}\int F_C\left(\frac{\tau}{L^z}\right) \exp{\left(-i2\pi fL^z \frac{\tau}{L^z}\right)} d\left(\frac{\tau}{L^z}\right).
\end{equation}
Corresponding to the reduced lag time $\tau/L^z$, we identify a reduced frequency $fL^z$. Then, the PSD follows a scaling ansatz
\begin{equation}
	S(f, L) \sim L^{z+1} F_S(fL^z),
	\label{eq_psd_1}
\end{equation}
where the scaling function $F_S(fL^z)$ in Eq.~(\ref{eq_psd_1}) is the Fourier transform of the scaling function $F_C(\cdot)$. After simplifying, we find
\begin{equation}
	F_S(u) \sim \begin{cases} {\rm constant},~~~~~~~~{\rm for}~ u \ll 1, \\ u^{-(1+1/z)}, ~~~~~~~ {\rm for}~ u \gg 1.   \end{cases}
	\label{eq_psd_2}
\end{equation}
It is easy to note [cf. Eqs.~(\ref{eq_psd_1})-(\ref{eq_psd_2})] that
\begin{equation}
	S(f, L) \sim \begin{cases} {\rm constant},~~~~~~~~{\rm for}~ f \ll L^{-z}, \\ f^{-(1+1/z)}, ~~~~~~~ {\rm for}~ f \gg L^{-z}.   \end{cases}
	\label{eq_psd_3}
\end{equation}

\begin{figure}[t]
	\centering
	\scalebox{1.0}{\includegraphics{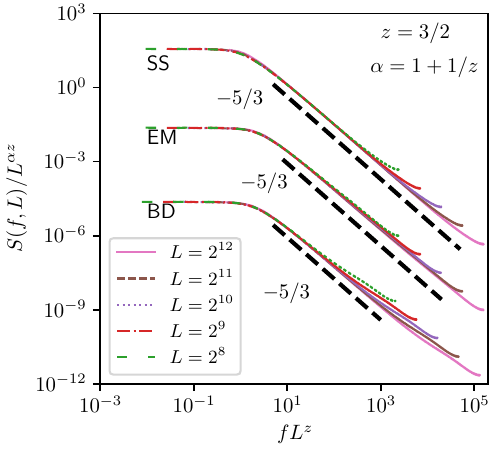}}
	\caption{The data collapse of the power spectrum $S(f, L)$ for the signal $\delta h_L(t)$ in different models. The BD and EM models belong to the KPZ universality class, and the results are consistent with the presented framework.}
	\label{fig_psd_all}
\end{figure}

Numerically, we directly compute the PSD of the temporal signal $\delta h_L(t)$ of length $T$ (the finite observation time) with the condition that $T \gg L^z$ as 
\begin{equation}
	\nonumber
	S(f, L) = \lim _{T\to \infty} \frac{1}{T}  \langle |\tilde{\delta h}_L(f)|^2\rangle,
\end{equation}
where the Fourier transform of the signal is  
\begin{equation}
\tilde{\delta h}_L\left(f = \frac{k}{T}\right) =  \sum_{t=0}^{T-1}\delta h_L(t) \exp\left( -i2\pi \frac{k}{T}t\right),\nonumber
\label{eq_ps_x1}
\end{equation}
with $k = 0, 1, 2, \cdots, T-1$. We don't normalize the PSD because the power in the low-frequency component shows an explicit system-size dependence. We consider a frequency window $f \in [1/T$, 1/2], with the lowest-frequency bins $\Delta f = 1/T$. Numerically, we implemented the standard Fast Fourier Transform (FFT) algorithm to compute the Fourier transform.

Figure \ref{fig_psd_0} (inset) presents the power spectra for the signal $\delta h_L(t)$ for different system sizes. We note that for a fixed system size, the power spectrum remains constant below a lower cutoff frequency (inverse of the correlation time $\mathcal{T}$) and shows a power law decaying behavior $\sim 1/f^{\alpha}$ in the nontrivial frequency regime  $L^{-z} \ll f \ll 1/2$. We also note that below the lower cutoff frequency power in low frequency component grows with the system size in a power law manner as $S(f\to 0, L) \sim L^{\alpha z}$ and the total power of the process scales as $P(L) \sim L^{(\alpha-1)z}$ [cf. Fig.~\ref{fig_psd_2}]. Numerically, we find that $\alpha z = 2.53(1)$ and $(\alpha-1)z = 1.02(1)$. It is easy to note that $z = 1.51(2)$. As shown in Fig.~\ref{fig_psd_0} (Main panel), applying FSS, we can obtain data collapse for the power spectra with the aid of the two exponents $\alpha z$ and $z$ by plotting scaled power $S(f, L)/L^{\alpha z}$ with the reduced frequency $fL^z$. A clean data collapse as obtained provides a precise estimate of the critical exponents.

Finally, to examine the extent of the scaling features observed for the KK model of the KPZ universality class, we also studied several discrete models (such as the EM, BD, and SS models) belonging to the same class. As shown in Fig.~\ref{fig_psd_all}, the data collapse of the power spectra confirms that the scaling features observed are not model-specific for the KPZ class.

\section{Correlation function scaling ansatz}{\label{sec_4}}
We assume that the temporal correlation function $C(\tau, \mathcal{T})$ of the stationary noisy signal $\delta h_L(t)$ that exhibits temporally scale-invariant features is a generalized homogeneous function of its arguments~\cite{Leonel05032026}, the lag-time $\tau$ and the correlation time $\mathcal{T}$. Then, we can write 
\begin{equation}
C(\tau, \mathcal{T}) \sim b^{-1}C(b^{y_\tau}\tau, b^{y_{\mathcal{T}}}\mathcal{T} ),
\label{eq_cr_function_1}
\end{equation}
where $b$ is a scaling factor.
Setting $b^{y_{\mathcal{T}}}\mathcal{T} = 1$ or $b \sim \mathcal{T}^{-1/y_{\mathcal{T}}}$ in Eq.~(\ref{eq_cr_function_1}), we get
\begin{equation}
C(\tau, \mathcal{T}) \sim \mathcal{T}^{1/y_{\mathcal{T}}}F_{C}(\tau/\mathcal{T}^{y_{\tau}/y_{\mathcal{T}}}).
\label{eq_cr_function_2}
\end{equation}
The argument of the scaling function should be dimensionless, suggesting $y_\tau = y_{\mathcal{T}}$. 
If we express the correlation function in its conventional form as 
\begin{equation}
C(\tau, \mathcal{T}) \sim \mathcal{T}^{\alpha-1}F_{C}(\tau/\mathcal{T}) \sim \tau^{\alpha-1}G_C(\tau/\mathcal{T}),
\label{eq_cr_function_3}
\end{equation}
where the scaling functions are $G_C(u) \sim F_C(u)/u^{\alpha-1}$ and $\alpha$ is the spectral exponnet. Comparing Eqs.~(\ref{eq_cr_function_2}) and (\ref{eq_cr_function_3}), we note that the correlation exponent is $\alpha-1 = 1/y_{\mathcal{T}}$.

In critical systems, the correlation time diverges with the linear extent of system size as $\mathcal{T}\sim L^z$, with dynamic exponent $z$. Writing Eq.~(\ref{eq_cr_function_3}) in terms of $L$, we get a scaling ansatz for the correlation function convenient for FSS as
\begin{equation}
C(\tau, L) \sim L^{z(\alpha-1)}F_{C}(\tau/L^z),
\label{eq_cr_function_4}
\end{equation}
where the scaling function becomes zero for $u \gg 1$. Demanding system size independent behavior in the nontrivial regime, we get
\begin{equation}
F_{C}(u) \sim u^{(\alpha-1)},~~~~{\rm for}~~u\ll 1,\nonumber
\label{eq_cr_function_5}
\end{equation}
The zero lag correlation or the variance varies as
\begin{equation}
C(0, L) \sim L^{z(\alpha-1)} \sim w^2(L) \sim L^{2\chi},\nonumber
\end{equation} 
suggesting $2\chi/z  = \alpha-1$.  
As the zero lag correlation function varies as $\sim L^{z(\alpha-1)}$, one can numerically estimate the critical exponents by examining system size scaling of the variance. 
Comparing Eqs.~(\ref{corr_scal_ansatz}) and (\ref{eq_cr_function_4}), we note that $z(\alpha-1) = 2\chi = 1$. Particularly, the spectral exponent is 
\begin{equation}
\alpha = 1+2\chi/z,
\label{eq_alpha_chi}
\end{equation} 
suggesting $\alpha = 5/3$ for the KPZ class as $\chi = 1/2$ and $z = 3/2$.

\section{Conclusion}{\label{sec_5}}
Related to the KPZ universality class, we studied four discrete systems (such as Kim-Kosterlitz, ballistic deposition, single-step, and etching models) of kinetic rough interfaces in one spatial dimension. In particular, we examine temporal correlations of height fluctuations. Notice that previous studies examined the PSD of the height noises and concluded that the underlying processes are non-stationary. Similarly, the two-time ensemble-averaged correlation functions reveal a signature of aging behavior for large systems. However, we evolve the dynamics for a long time to attain a stationary state, ensuring that the finite observation time remains much larger than the underlying correlation time, which scales as $\mathcal{T} \sim L^z$. We compute the two-time autocorrelation function, which varies nonexponentially $\tau^{1/z}$ for $\tau \ll T^{z}$ and becomes zero thereafter. We can obtain a data collapse for the correlation functions for different system sizes using FSS. We also compute the PSD that displays a power law scaling $1/f^{\alpha}$ with the spectral exponent value $\alpha = 1+1/z = 5/3$ [cf. Eq.~(\ref{eq_alpha_chi})] in the nontrivial frequency regime $L^{-z}\ll f\ll 1/2$. The power shows a lower cutoff frequency, below which it remains constant. Our scaling arguments can relate the two characterizations. Our main conclusion is that for a small system, it is possible to reach a stationary state, and the Wiener-Khinchin theorem remains valid. To examine the extent of stationary state fluctuations, it would be interesting to investigate the role of dimensionality and anomalous rough interfaces. Our work clarifies the experimental limitation due to the diverging feature of the correlation time, which is why the lower cutoff of the power spectrum remains typically inaccessible.

\section{Acknowledegment}
RC thanks UGC, India, for providing the funding through a Senior Research Fellowship. We also acknowledge IIT BHU, Varanasi, for the supercomputing facility (PARAM Shivay supercomputer).

\bibliography{s1sources}
\bibliographystyle{myrev}

\end{document}